\documentclass[11pt,a4paper]{article}

\pdfoutput=1

\usepackage{amsmath}
\usepackage{amssymb}
\usepackage[square,comma,numbers,sort&compress]{natbib}
\usepackage[retainorgcmds]{IEEEtrantools}
\usepackage{tikz}
\usepackage{titling}
\usepackage{authblk}
\usepackage[colorlinks,linkcolor=blue,citecolor=blue,urlcolor=blue]{hyperref}
\usepackage{fullpage}
\usepackage{quant_defs}

\usetikzlibrary{shapes.misc,shapes.symbols,shapes.geometric,arrows,
  plotmarks,decorations.pathmorphing}
\tikzset{gridlines/.style={very thin,color=gray!50},
  every picture/.append style={scale=1}}

\definecolor{darkgreen}{rgb}{0,0.5,0}

\pretitle{\begin{center}\Large\sffamily\bf}
\posttitle{\par\end{center}}

\newcommand{\psx}[0]{\psi_{00}}
\newcommand{\psy}[0]{\psi_{01}}
\newcommand{\psu}[0]{\psi_{10}}
\newcommand{\psv}[0]{\psi_{11}}

\newcommand{\psX}[0]{\Psi_{00}}
\newcommand{\psY}[0]{\Psi_{01}}
\newcommand{\psU}[0]{\Psi_{10}}
\newcommand{\psV}[0]{\Psi_{11}}

\newcommand{\phx}[0]{\phi_{00}}
\newcommand{\phy}[0]{\phi_{01}}
\newcommand{\phu}[0]{\phi_{10}}
\newcommand{\phv}[0]{\phi_{11}}

\newcommand{\etal}[0]{\emph{et al.}}

\begin{document}

\title{Effects of preparation and measurement misalignments on \\ the security
  of the BB84 quantum key distribution protocol}
\date{20 March 2013}
\author[1]{Erik Woodhead\thanks{Erik.Woodhead@ulb.ac.be}}
\author[1]{Stefano Pironio}
\affil[1]{Laboratoire d'Information Quantique, Universit\'e Libre de
  Bruxelles, Belgium}

\maketitle

\begin{abstract}
  The ideal Bennett-Brassard 1984 (BB84) quantum key distribution protocol is
  based on the preparation and measurement of qubits in two alternative bases
  differing by an angle of $\pi/2$. Any real implementation of the protocol,
  though, will inevitably introduce misalignments in the preparation of the
  states and in the alignment of the measurement bases with respect to this
  ideal situation. Various security proofs take into account (at least
  partially) such errors, i.e., show how Alice and Bob can still distil a
  secure key in the presence of these imperfections. Here, we consider the
  complementary problem: how can Eve exploit misalignments to obtain more
  information about the key than would be possible in an ideal
  implementation? Specifically, we investigate the effects of misalignment
  errors on the security of the BB84 protocol in the case of individual
  attacks, where necessary and sufficient conditions for security are
  known. Though the effects of these errors are small for expected deviations
  from the perfect situation, our results nevertheless show that Alice and
  Bob can incorrectly conclude that they have established a secure key if the
  inevitable experimental errors in the state preparation and in the
  alignment of the measurements are not taken into account. This gives
  further weight to the idea that the formulation and security analysis of
  any quantum cryptography protocol should be based on realistic assumptions
  about the properties of the apparatus used. Additionally, we note that BB84
  seems more robust against alignment imperfections if both the $x$ and $z$
  bases are used to generate the key.
\end{abstract}


\section{Introduction\label{sec:intro}}

The use of quantum systems to accomplish cryptographic tasks promises levels
of security unachievable with any classical system. With these benefits,
however, comes an added difficulty. Unlike classical protocols intended for
execution on a digital computing device and whose security is purely based on
the \emph{mathematical} properties of the device's outputs, quantum protocols
make use of analogue systems and their security is intrinsically
\emph{physical}: it depends on the fact that device's output was obtained by
measuring, e.g., the polarisation of a single photon along well defined
orientations. Deviations from the ideal situation, which are an
all-or-nothing affair in a digital algorithm and can typically be eliminated
with some very large probability, therefore become inevitable to some degree
in quantum protocols.

The Bennett-Brassard 1984 (BB84) protocol \cite{ref:bb84} for quantum key
distribution \cite{ref:gr02,ref:sbp09}, for instance, requires that one party
(`Alice') prepares and sends a sequence of random qubits taken from the set
$\{\ket{\psi_{bm}}\}$, where the indices $b, m \in \{0,1\}$ can be
interpreted as a choice of basis and bit, respectively. The other party
(`Bob') then randomly measures each qubit he receives in one of two bases
$\{\ket{\phx}, \ket{\phy}\}$ or $\{\ket{\phu}, \ket{\phv}\}$. In its ideal
formulation, the states $\{\ket{\psi_{b0}}, \ket{\psi_{b1}}\}$ prepared by
Alice are supposed to form a basis and therefore to be orthogonal,
\begin{equation}
  \label{eq:ortho}
  \braket{\psi_{b0}}{\psi_{b1}} = 0 \quad \text{ for } b = 0, 1 \,.
\end{equation} 
Furthermore, the two bases on Alice's and on Bob's sides are supposed to
differ exactly by an angle of $\pi/2$, i.e., to satisfy the
relations\footnote{In addition, in the ideal formulation of the BB84
  protocol, the bases on Alice's and Bob's sides are usually taken to be
  perfectly aligned, i.e., $\ket{\psi_{bm}}=\ket{\phi_{bm}}$. But any
  misalignment between the two bases can always be absorbed in the unitary
  transformation performed by Eve on the states emitted by Alice and thus
  has no incidence on the security of the protocol.}
\begin{subequations}
  \label{eq:bb84_states}
  \begin{IEEEeqnarray}{rCl}
    \ket{\psu} &=& \tfrac{1}{\sqrt{2}} \bigl[ \ket{\psx}
    + \ket{\psy} \bigr] \,, \\
    \ket{\psv} &=& \tfrac{1}{\sqrt{2}} \bigl[ \ket{\psy}
    - \ket{\psx} \bigr] \,,
  \end{IEEEeqnarray}
\end{subequations}
and
\begin{subequations}
  \label{eq:bb84_states_b}
  \begin{IEEEeqnarray}{rCl}
    \ket{\phu} &=& \tfrac{1}{\sqrt{2}} \bigl[ \ket{\phx}
    + \ket{\phy} \bigr] \,, \\
    \ket{\phv} &=& \tfrac{1}{\sqrt{2}} \bigl[ \ket{\phy}
    - \ket{\phx} \bigr] \,.
  \end{IEEEeqnarray}
\end{subequations}
While existing security proofs for BB84 can deal with an arbitrary noise in
the quantum channel from Alice to Bob, they usually assume that the states
prepared by Alice and that the measurements performed by Bob satisfy
precisely the conditions \eqref{eq:ortho}, \eqref{eq:bb84_states}, and
\eqref{eq:bb84_states_b}. In a realistic execution of the protocol, however,
experimental errors are inevitable. For instance, the measurement of a
polarisation qubit cannot be more precise than $2^{\circ}$ or $4^{\circ}$ (on
the Bloch sphere) due to the intrinsic uncertainty of the polarisation
rotator used. Such imperfections may allow an eavesdropper to gain more
information about the shared key than existing security proofs would imply.

Here we illustrate the effects that imperfections in the preparation of the
states and in the alignment of the measurement bases could have on the
performance of quantum cryptography protocols, using the BB84 protocol as our
example.

We note that proofs of security of BB84 have been proposed that relax
conditions \eqref{eq:ortho} and \eqref{eq:bb84_states}
\cite{ref:kp03,ref:k09}, conditions \eqref{eq:bb84_states_b} \cite{ref:my98},
conditions \eqref{eq:bb84_states} and \eqref{eq:bb84_states_b}
\cite{ref:tr11}, and also that take into account certain particular
modifications of all three conditions \eqref{eq:ortho},
\eqref{eq:bb84_states}, and \eqref{eq:bb84_states_b} in the context of
collective attacks \cite{ref:gl04}. A proof of security in the asymptotic
limit valid against arbitrary deviations from the three conditions
\eqref{eq:ortho}, \eqref{eq:bb84_states}, and \eqref{eq:bb84_states_b} has
also been reported in \cite{ref:mls10}. These types of analyses, however, are
not routinely considered and scarcely used in practical implementations of
BB84 \cite{ref:g11}. The main objective of this paper is to draw attention to
this issue.

Rather than deriving a new security proof, our aim is to demonstrate an
explicit advantage gained by an eavesdropper. We therefore restrict our
analysis to individual attacks -- where contrarily to more general types of
attacks, necessary and sufficient conditions for security are known -- and
optimise over all possible attacks of this type in the presence of
imperfections. We emphasise that though security proofs against more general
types of attacks, such as those mentioned above, do report keyrates that are
lower than in the ideal case, it is not a priori clear that these observed
reductions in security are genuine and not an artefact of a suboptimal
security proof. In the case of individual attacks, however, optimal criteria
for security are known, and thus any reduction in the keyrate that we observe
illustrates some genuine advantage gained by the eavesdropper. Furthermore,
general security proofs bound security ``from below'', ruling out possible
successful attacks by an eavesdropper below a certain threshold. In
optimising explicitly over individual attacks, we bound security ``from
above''. Our results can thus also be viewed as representing an \emph{upper}
bound on security: we strictly prove that non-ideal BB84 implementations of
the type we consider are \emph{insecure} above a certain threshold.

For simplicity, we consider the case where the states emitted by
Alice still form two orthonormal bases as in \eqref{eq:ortho}. (Any deviation
from \eqref{eq:ortho} can only reinforce the effects of imperfections that we
illustrate here.) We suppose, however, that Alice's preparation and Bob's
measurement bases are not exactly mutually unbiased, but that they differ by
angles $\alpha$ and $\beta$, respectively, different from $\pi/2$. That is,
we suppose instead of \eqref{eq:bb84_states} and \eqref{eq:bb84_states_b}
that
\begin{subequations}
  \label{eq:bb84d_alice}
  \begin{IEEEeqnarray}{rCl}
    \ket{\psu} &=& \cos\bigl(\tfrac{\alpha}{2}\bigr) \ket{\psx}
    + \sin\bigl(\tfrac{\alpha}{2}\bigr) \ket{\psy} \,, \\
    \ket{\psv} &=& \cos\bigl(\tfrac{\alpha}{2}\bigr) \ket{\psy}
    - \sin\bigl(\tfrac{\alpha}{2}\bigr) \ket{\psx} \,,
  \end{IEEEeqnarray}
\end{subequations}
and 
\begin{subequations}
  \label{eq:bb84d_bob}
  \begin{IEEEeqnarray}{rCl}
    \ket{\phu} &=& \cos\bigl(\tfrac{\beta}{2}\bigr) \ket{\phx}
    + \sin\bigl(\tfrac{\beta}{2}\bigr) \ket{\phy} \,, \\
    \ket{\phv} &=& \cos\bigl(\tfrac{\beta}{2}\bigr) \ket{\phy}
    - \sin\bigl(\tfrac{\beta}{2}\bigr) \ket{\phx} \,.
  \end{IEEEeqnarray}
\end{subequations}
It is clear that such errors will in general reduce the security of BB84. For
example, in the extreme case where the two bases accidentally coincide
($\alpha, \beta = 0$), an eavesdropper could perfectly clone the states sent
by Alice without revealing her presence. Using a combination of analytical
techniques and numerical optimisation, we demonstrate here more generally a
reduction in the extractable secret keyrate of the BB84 protocol against
individual attacks, for a given quantum bit error rate (QBER), when $\alpha,
\beta \neq \pi/2$.

Though the reduction in the keyrate that we observe is small for deviations
from the ideal situation expected in realistic implementations, our results
nevertheless show that Alice and Bob can erroneously conclude that they have
established a secure key if the inevitable experimental errors in the
alignment of the bases are not taken into account. Though our findings are
restricted to individual attacks, it is reasonable to expect that similar
results hold in full generality. This gives further weight to the idea that
the formulation and security analysis of any quantum cryptography protocol
should be based on realistic assumptions about the properties of the
apparatus used.

This conclusion goes in a similar direction as that which can be drawn from
the recent weaknesses discovered in certain QKD implementations, such as
\cite{ref:xql10,ref:lw10}. Note though that our work has a very different
perspective. Indeed, contrary to \cite{ref:xql10,ref:lw10}, our results do
not uncover an implementation flaw in an otherwise theoretically secure
scheme -- a flaw which could therefore be fixed purely at the implementation
level. The message that we want to convey here is rather that in any trusted
and ``secure'' QKD implementation, uncertainties in the preparation of the
quantum states and in the alignment of the measurement bases will inevitably
be present and may affect the security. These uncertainties must therefore be
accounted for at a \emph{theoretical} level either by adapting the security
proof or by moving to device-independent \cite{ref:my04,ref:ab07} or
semi-device-independent schemes \cite{ref:pb11,ref:wlp13,ref:lcq12}.

The present work originates from a loose collaboration with the authors of
\cite{ref:bg11,ref:rfs12}, who along similar lines have explored the effect
of imperfections in the alignment of measurement bases on the
characterisation of quantum resources through quantum state tomography and
entanglement witnesses.

Our results are presented in more detail in section~\ref{sec:results};
technical details are deferred to section~\ref{sec:technical}.

\section{Results\label{sec:results}}

\subsection{Problem definition}

We begin by briefly recounting the BB84 protocol. As recalled above, one
party (Alice) prepares random qubits from the set $\{\ket{\psi_{bm}}\}$, and
transmits them to a second party (Bob). Bob then measures each qubit that he
receives in one of two bases $\{\ket{\phi_{b0}}, \ket{\phi_{b1}}\}$, randomly
choosing between $b=0$ and $b=1$ each time, and stores the results. After
discarding the cases where the choices of basis do not match, Alice and Bob
share a so-called ``sifted key'', with Bob's version of the key likely
containing errors compared with Alice's. By sacrificing a part of the sifted
key, Alice and Bob can estimate the quantum bit error rate (QBER) $Q$, which
is defined in terms of the observed coincidence rates $p^{(b)}(m,n)$ of Alice
sending a state encoding bit $m$ and Bob measuring $n$, given basis
$b$. Assuming that the QBER is the same in both bases, it can be defined as
\begin{equation}
  Q = \tfrac{1}{2} \sum_{b \in \{0,1\}}
  \bigl( p^{(b)}(0,1) + p^{(b)}(1,0) \bigr) \,.
\end{equation}
Following this, error correction and privacy amplification are applied. In
the case of one-way communication from Alice to Bob, the asymptotic keyrate
secure against individual attacks is given by the Csisz\'ar-K\"orner bound
\cite{ref:ck78}:
\begin{equation}
  \label{eq:ck_bound}
  r = I(A:B) - I(A:E) \,,
\end{equation}
where $I(A:B)$ denotes the mutual information between Alice and Bob and
$I(A:E)$ between Alice and Eve. We recall that, in the case of individual
attacks, Eve performs the same unitary attack on each of Alice's qubits, but
is allowed to possess a quantum memory and can delay her measurements on the
states in her possession until after the bases are revealed. Fuchs \etal{}\
show in \cite{ref:fg97} that the highest secure asymptotic keyrate under
conditions \eqref{eq:ortho}, \eqref{eq:bb84_states}, \eqref{eq:bb84_states_b}
is given in terms of $Q$ by
\begin{equation}
  \label{eq:bb84_keyrate}
  r = h \bigl(\tfrac{1}{2} - \sqrt{Q(1-Q)}\bigr) - h(Q) \,,
\end{equation}
where $h$ is the binary entropy function.

Our task is to minimise the expression \eqref{eq:ck_bound} for a given QBER
$Q$ using the preparation and measurement bases defined by
\eqref{eq:bb84d_alice} and \eqref{eq:bb84d_bob} rather than the ideal
ones. To simplify the analysis we will assume that the errors observed
between Alice and Bob are symmetric, i.e.\
\begin{equation}
  \label{eq:symerror}
  p^{(0)}(0,1) = p^{(0)}(1,0) = p^{(1)}(0,1) = p^{(1)}(1,0) \,.
\end{equation}
Given our assumptions about the symmetries in the errors observed by Alice
and Bob, $I(A:B)$ is a simple function of $Q$:
\begin{equation}
  I(A:B) = 2 - h(Q) \,.
\end{equation}

In general there need not be such symmetries in the joint probabilities
$p^{(b)}_{\mathrm{AE}}(m,q)$ shared between Alice and Eve, and $I(A:E)$ is
accordingly more complicated. In each basis it will be convenient to
parameterise these quantities in terms of an error $Q^{(b)}_{\mathrm{AE}}$
analogous to the QBER, and an offset $\delta^{(b)}$:
\begin{subequations}
  \label{eq:probs_q_delta}
  \begin{IEEEeqnarray}{rCl}
    p_{\mathrm{AE}}^{(b)}(0,0)
    &=& \tfrac{1}{2}(1 - Q^{(b)}_{\mathrm{AE}} - \delta^{(b)}) \,, \\
    p_{\mathrm{AE}}^{(b)}(0,1)
    &=& \tfrac{1}{2}(Q^{(b)}_{\mathrm{AE}} + \delta^{(b)}) \,, \\
    p_{\mathrm{AE}}^{(b)}(1,0)
    &=& \tfrac{1}{2}(Q^{(b)}_{\mathrm{AE}} - \delta^{(b)}) \,, \\
    p_{\mathrm{AE}}^{(b)}(1,1)
    &=& \tfrac{1}{2}(1 - Q^{(b)}_{\mathrm{AE}} + \delta^{(b)}) \,.
  \end{IEEEeqnarray}
\end{subequations}
The inverse relations are $Q_{\mathrm{AE}}^{(b)} = p_{\mathrm{AE}}^{(b)}(0,1)
+ p_{\mathrm{AE}}^{(b)}(1,0)$ and $\delta^{(b)} = p_{\mathrm{AE}}^{(b)}(0,1)
- p_{\mathrm{AE}}^{(b)}(1,0)$. The mutual information between Alice and Eve
is given by
\begin{equation}
  I(A:E) = 1 + \tfrac{1}{2} \bigl( I^{(0)}(A:E) + I^{(1)}(A:E) \bigr) \,,
\end{equation}
where $I^{(b)}(A:E)$ is the mutual information in a single basis, determined
by the joint probabilities $p^{(b)}_{\mathrm{AE}}(m,n)$.

We present results for the numerical optimisation of this problem in the next
subsection. Details of the parameterisation and techniques employed are
deferred to section~\ref{sec:technical}.
 
\subsection{Optimisation results\label{sec:opt_results}}

In numerically evaluating the keyrate, it generally seems to be the case, as
one might expect, that the minimal keyrate is found for a unitary interaction
that gives Eve symmetric information about the bits in Alice's possession. In
terms of the parameterisation introduced at the end of the previous section,
this is the case where $\delta^{(0)} = \delta^{(1)} = 0$ and
$Q^{(0)}_{\mathrm{AE}} = Q^{(1)}_{\mathrm{AE}} \equiv Q_{\mathrm{AE}}$. The
keyrate is then a simple function of $Q$ and $Q_{\mathrm{AE}}$:
\begin{equation}
  r = h(Q_{\mathrm{AE}}) - h(Q) \,.
\end{equation}
Supported by a few test cases, this simplification was applied in the results
we now present. (Note that even if Eve's optimal attack does not generally
satisfy this symmetry, our results still represent an upper bound on the
secure keyrate, which conclusively shows that Eve can gain information by
exploiting preparation and measurement imperfections with respect to the
ideal case.)

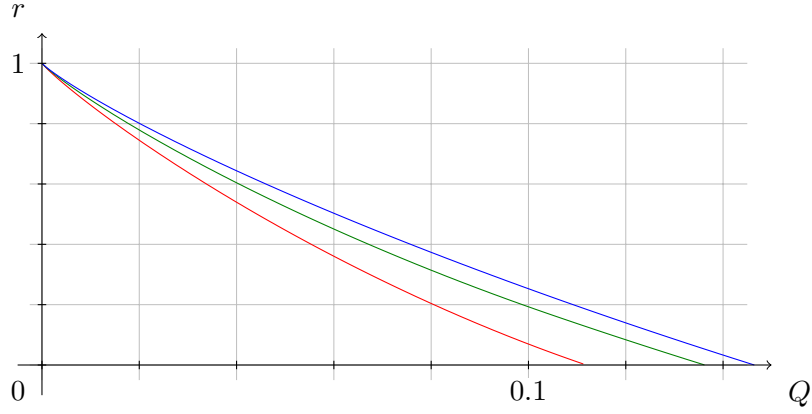
\begin{figure}[htbp]
  \centering
  \begin{tikzpicture}[xscale=64,yscale=4,inner sep=1pt]
    \draw[gridlines,xstep=0.02,ystep=0.2] (-0.0025,-0.05) grid (0.145,1.05);

    \draw[smooth,color=red] plot file{plots/qber_r_70.table};
    \draw[smooth,color=darkgreen] plot file{plots/qber_r_80.table};
    \draw[smooth,color=blue] plot file{plots/qber_r_90.table};

    \draw[->] (-0.005,0) -- (0.15,0) node[below right=5pt,fill=white]{$Q$};
    \draw[->] (0,-0.1) -- (0,1.1) node[above left=5pt,fill=white]{$r$};

    \foreach \x in {0,0.02,...,0.16001}
      \draw (\x, -0.4pt) -- (\x, 0.4pt);

    \foreach \x in {0,0.2,...,1.001}
      \draw (-0.025pt, \x) -- (0.025pt, \x);

    \draw (0,0) node[below left=5pt,fill=white]{$0$};
    \draw (0.1,0) node [below=5pt,fill=white]{$0.1$};
    \draw (0,1) node[left=5pt,fill=white]{$1$};
  \end{tikzpicture}
  \caption{Variation of keyrate with QBER for $\theta =
    \textcolor{blue}{90^{\circ}}$, $\textcolor{darkgreen}{80^{\circ}}$, and
    $\textcolor{red}{70^{\circ}}$, corresponding to the worst-case scenarios
    for errors of $0^{\circ}$, $5^{\circ}$, and $10^{\circ}$ respectively.}
  \label{fig:keyrate_qber}
\end{figure}

Figure~\ref{fig:keyrate_qber} is a plot of the optimised keyrate as a
function of $Q$ for a few fixed values of $\alpha = \beta = \theta$. The
values of $\theta$ used are $90^{\circ}$ (the ideal case), $80^{\circ}$, and
$70^{\circ}$. The latter two are the worst-case scenarios if there are
absolute experimental errors of respectively $5^{\circ}$ and $10^{\circ}$ on
the orientations of the bases both used by Alice and measured by Bob. That
is, if Alice and Bob know, say, that their devices are accurate to within
five degrees, i.e., $80^{\circ} \leq \alpha, \beta \leq 90^{\circ}$, then the
worst keyrate that we have found corresponds to the situation
$\alpha=\beta=\theta=80^{\circ}$. The worst-case scenario is thus that the
largest possible error on the orientation of the devices is systematic.

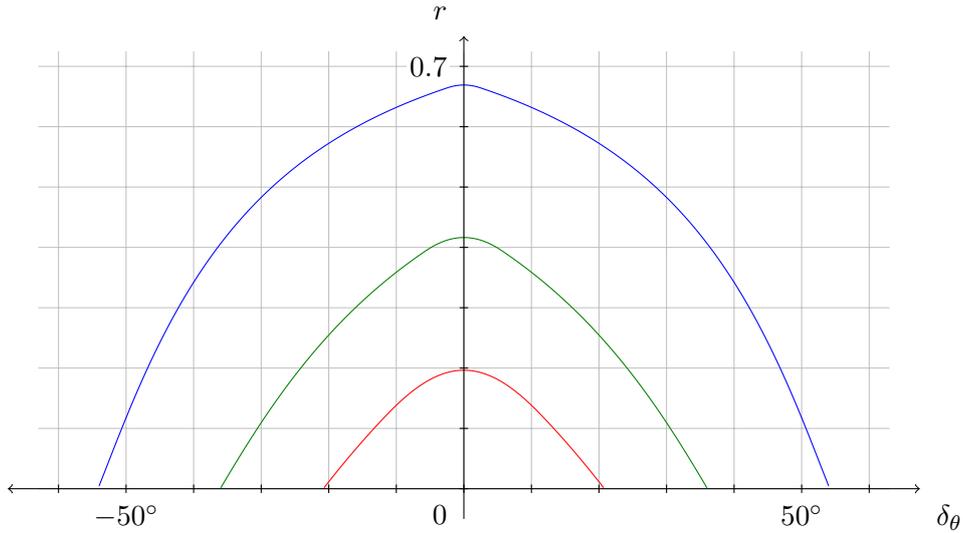
\begin{figure}[htbp]
  \centering
  \begin{tikzpicture}[xscale=8,yscale=8,inner sep=1pt]
    \draw[gridlines,xstep=0.11111111,ystep=0.1] (-0.7,0) grid (0.7,0.725);

    \draw[smooth,color=red] plot file{plots/theta_r_075.table};
    \draw[smooth,color=darkgreen] plot file{plots/theta_r_05.table};
    \draw[smooth,color=blue] plot file{plots/theta_r_025.table};

    \draw[<->] (-0.75,0) -- (0.75,0)
      node[below right=5pt,fill=white]{$\delta_{\theta}$};
    \draw[->] (0,-0.05) -- (0,0.75) node[above left=5pt,fill=white]{$r$};

    \foreach \x in {-0.66666667,-0.55555556,...,0.67}
      \draw (\x, -0.2pt) -- (\x, 0.2pt);

    \foreach \x in {0,0.1,...,0.7001}
      \draw (-0.2pt, \x) -- (0.2pt, \x);


    \draw (0,0) node[below left=5pt,fill=white]{$0$};
    \draw (-0.55555556,0) node [below=5pt,fill=white]{$-50^{\circ}$};
    \draw (0.55555556,0) node[below=5pt,fill=white]{$50^{\circ}$};
    \draw (0,0.7) node[left=5pt,fill=white]{$0.7$};
  \end{tikzpicture}
  \caption{Variation of keyrate with angle $\delta_{\theta} = 90^{\circ} -
    \theta$, for $Q = \textcolor{blue}{\tfrac{1}{4}} Q_{0}$,
    $\textcolor{darkgreen}{\tfrac{1}{2}} Q_{0}$, and
    $\textcolor{red}{\tfrac{3}{4}} Q_{0}$, where $Q_{0} \approx 0.1464$ is
    the upper secure bound on the QBER.}
  \label{fig:keyrate_theta}
\end{figure}

Figure~\ref{fig:keyrate_theta} is a plot of the minimised keyrate as a
function of the deviation $\delta_{\theta} = \pi/2 - \theta$ from the ideal
case, for QBERs of $\tfrac{1}{4} Q_{0}$, $\tfrac{1}{2} Q_{0}$, and
$\tfrac{3}{4} Q_{0}$, where $Q_{0} = \frac{1}{2} - \frac{1}{4} \sqrt{2}
\approx 0.1464$ is the maximum tolerable QBER in the ideal case.

\begin{figure}[htbp]
  \centering
  \begin{tikzpicture}[xscale=5,yscale=32,inner sep=1pt]
    \draw[gridlines,xstep=0.111111111,ystep=0.02] (-1.05,0)
      grid (1.05,0.165);

    \draw[dashed,color=red] (-1.08, 0.11002786) -- (1.08, 0.11002786)
      node [above right=3pt,color=red,fill=white] {S-P};

    \draw[smooth,color=blue] plot file{plots/secure_q.table};

    \draw[->] (-1.12,0) -- (1.12,0)
      node[below right=5pt,fill=white]{$\delta_{\theta}$};
    \draw[->] (0,-0.02) -- (0,0.17) node[above left=5pt,fill=white]{$Q$};

    \foreach \x in {-1.0,-0.888888889,...,1.001}
      \draw (\x, -0.05pt) -- (\x, 0.05pt);

    \foreach \x in {0,0.02,...,0.1601}
      \draw (-0.2pt, \x) -- (0.2pt, \x);

    \draw (0,0) node[below left=5pt,fill=white]{$0$};
    \draw (-1,0) node[below=5pt,fill=white]{$-90^{\circ}$};
    \draw (1,0) node[below=5pt,fill=white]{$90^{\circ}$};
    \draw (0,0.1) node[left=5pt,fill=white]{$0.1$};
  \end{tikzpicture}
  \caption{Maximum secure QBER as a function of $\delta_{\theta} = 90^{\circ}
    - \theta$. The horizontal dashed line (\textcolor{red}{S-P}) corresponds
    to the Shor-Preskill bound of about 0.11.}
  \label{fig:secure_qber}
\end{figure}
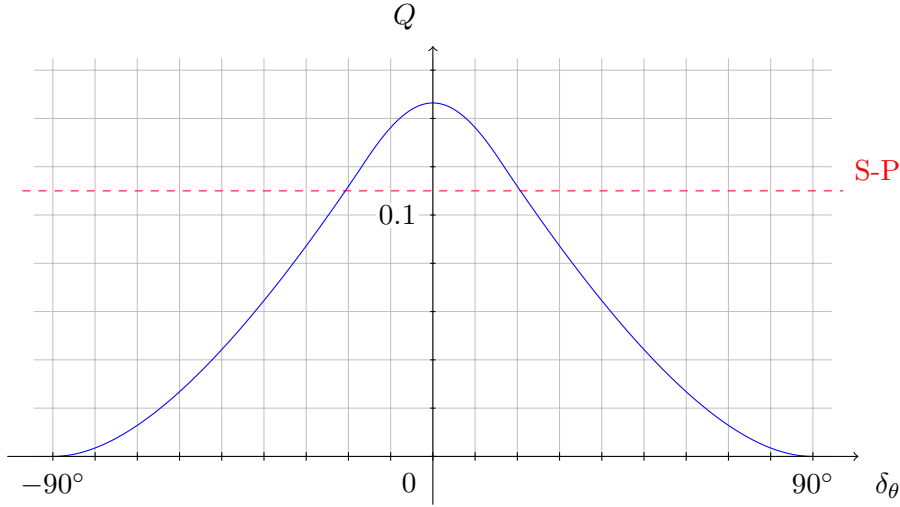

Finally, figure~\ref{fig:secure_qber} is a plot of the upper secure bound on the
QBER as a function of the deviation $\delta_{\theta} = \pi/2 - \theta$. The
Shor-Preskill bound of 0.11 \cite{ref:sp00}, representing the best known
threshold QBER below which an ideal BB84 implementation is known to be
secure against arbitrary attacks, is added for comparison.

\subsection{Discussion\label{sec:discussion}}

Assuming that Alice and Bob observe errors that are symmetric, according to
\eqref{eq:symerror}, using a combination of analytical and numerical
techniques we have determined upper bounds on the keyrate for preparation and
measurement devices characterised by the misalignment angles $\alpha$ and
$\beta$ defined in \eqref{eq:bb84d_alice} and \eqref{eq:bb84d_bob}. As soon
as $\alpha, \beta \neq \pi/2$, we find that these upper bounds are lower than
the optimal keyrate \eqref{eq:bb84_keyrate} for a given QBER, therefore
showing that imperfections in the preparation and measurement devices can be
exploited by an eavesdropper if they are not taken into account in the
security proof. We also draw attention to the fact that the threshold QBER
illustrated in figure~\ref{fig:secure_qber} drops below the Shor-Preskill
bound of about 0.11 for deviation angles larger than about $20.7^{\circ}$,
demonstrating that the Shor-Preskill keyrate is certainly insecure in this
case.

The upper bounds that we have obtained correspond to the best individual
attack that is symmetric, i.e., that satisfies $\delta^{(0)} = \delta^{(1)} =
0$ and $Q^{(0)}_{\mathrm{AE}} = Q^{(1)}_{\mathrm{AE}} \equiv
Q_{\mathrm{AE}}$. We have numerically verified in a few test cases that the
best overall individual attack satisfies this symmetry condition. We thus
expect our upper bounds on the keyrate to actually correspond to the
\emph{optimal} keyrates in the presence of imperfections of the type we
consider.

If Alice and Bob know that their devices are accurate to within a given
precision $\delta_{\theta}$, they should assume, for the purpose of proving
security, that their devices are characterised by the angles $\alpha$ and
$\beta$ compatible with this precision that yield the worst-case keyrate. We
verified in a few test cases that this happens for the smallest angles
$\alpha$ and $\beta$ consistent with the set error, at least in the case
where the set error is the same on Alice's and Bob's devices. It is for this
reason that the above figures are plotted for values of the angles satisfying
$\alpha = \beta = \theta = \pi/2 - \delta_{\theta}$.

All the results that we have presented here were obtained for the case where
both bases are used to establish the secret key. One may also consider the
variant of BB84 in which only one basis is used to generate the key
\cite{ref:lca05}. In the ideal case, this results in a keyrate that is
asymptotically twice as high, as the sifting step, where half of the results
are discarded, is no longer necessary. We have also adapted our analysis to
this situation and have found that for high QBERs the two-basis protocol
results in a higher keyrate than the single-basis one, suggesting that the
former is more robust against alignment errors.

Finally, we remind the reader that throughout our analysis, we have assumed
that the states prepared by Alice define a basis, i.e., satisfy
\eqref{eq:ortho}. Relaxing this condition could only strengthen the effects
of imperfections observed here.

\section{Technical details\label{sec:technical}}
\subsection{Eve's interaction\label{sec:eve_probe}}

The model applied here is a straightforward adaptation of the one considered
in \cite{ref:fg97}. In the worst-case scenario the eavesdropper (Eve) has
replaced the quantum channel between Alice and Bob with a lossless channel,
before appending an ancilla to the state sent by Alice and applying a unitary
operation with the intent of cloning the communication. We express the
interaction as
\begin{subequations}
  \begin{IEEEeqnarray}{rCl}
    \ket{\psx} \ket{0} &\mapsto& \ket{\psX} \,, \\
    \ket{\psy} \ket{0} &\mapsto& \ket{\psY} \,,
  \end{IEEEeqnarray}
\end{subequations}
in the basis $b = 0$, and similarly
\begin{subequations}
  \begin{IEEEeqnarray}{rCl}
    \ket{\psu} \ket{0} &\mapsto& \ket{\psU} \,, \\
    \ket{\psv} \ket{0} &\mapsto& \ket{\psV} \,,
  \end{IEEEeqnarray}
\end{subequations}
in the basis $b = 1$, where the states $\{\ket{\Psi_{bx}}\}$ are states in
the Hilbert space $\mathcal{H}_{\mathrm{B}} \otimes \mathcal{H}_{\mathrm{E}}$
accessible to Bob and Eve. Linearity of the unitary interaction implies that
these states obey the same relations as $\{\ket{\psi_{bx}}\}$. Specifically,
\begin{subequations}
  \label{eq:probe_transf}
  \begin{IEEEeqnarray}{rCl}
    \ket{\psU} &=& \cos\bigl(\tfrac{\alpha}{2}\bigr) \ket{\psX}
    + \sin\bigl(\tfrac{\alpha}{2}\bigr) \ket{\psY} \,, \\
    \ket{\psV} &=& \cos\bigl(\tfrac{\alpha}{2}\bigr) \ket{\psY}
    - \sin\bigl(\tfrac{\alpha}{2}\bigr) \ket{\psX} \,.
  \end{IEEEeqnarray}
\end{subequations}
In order to parameterise the interaction, we set
\begin{subequations}
  \label{eq:probe_xy}
  \begin{IEEEeqnarray}{rCl}
    \ket{\psX} &=& \ket{\phx} \bigl( \ket{a} + \ket{b} \bigr)
    + \ket{\phy} \bigl( \ket{c} + \ket{d} \bigr) \,, \\
    \ket{\psY} &=& \ket{\phy} \bigl( \ket{a} - \ket{b} \bigr)
    + \ket{\phx} \bigl( \ket{c} - \ket{d} \bigr) \,,
  \end{IEEEeqnarray}
\end{subequations}
and
\begin{subequations}
  \label{eq:probe_uv}
  \begin{IEEEeqnarray}{rCl}
    \ket{\psU} &=& \ket{\phu} \bigl( \ket{a'} + \ket{b'} \bigr)
    + \ket{\phv} \bigl( \ket{c'} + \ket{d'} \bigr) \,, \\
    \ket{\psV} &=& \ket{\phv} \bigl( \ket{a'} - \ket{b'} \bigr)
    + \ket{\phu} \bigl( \ket{c'} - \ket{d'} \bigr) \,,
  \end{IEEEeqnarray}
\end{subequations}
where $\ket{a}, \ket{b}, \ket{c}, \ket{d} \in \mathcal{H}_{\mathrm{E}}$ are
(not necessarily normalised) states accessible to Eve whose ``metric''
$\gamma_{ij} = \braket{i}{j},\, i,j \in \{a,b,c,d\}$ completely defines Eve's
interaction. Combining \eqref{eq:probe_xy} and \eqref{eq:probe_uv} with
\eqref{eq:probe_transf} and \eqref{eq:bb84d_bob}, we extract the relations
\begin{subequations}
  \begin{IEEEeqnarray}{rCl}
    \ket{a'} &=& \cos(\Delta) \ket{a} + \sin(\Delta) \ket{d} \,, \\
    \ket{d'} &=& \cos(\Delta) \ket{d} - \sin(\Delta) \ket{a} \,,
  \end{IEEEeqnarray}
\end{subequations}
and
\begin{subequations}
  \begin{IEEEeqnarray}{rCl}
    \ket{b'} &=& \cos(\theta) \ket{b} + \sin(\theta) \ket{c} \,, \\
    \ket{c'} &=& \cos(\theta) \ket{c} - \sin(\theta) \ket{b} \,,
  \end{IEEEeqnarray}
\end{subequations}
where we have set
\begin{subequations}
  \label{eq:defs_delta_theta}
  \begin{IEEEeqnarray}{rCl}
    \Delta &=& \frac{\beta - \alpha}{2} \,, \\
    \theta &=& \frac{\beta + \alpha}{2} \,.
  \end{IEEEeqnarray}
\end{subequations}
The problem now is to identify the metric $\gamma_{ij}$ which will maximise
the information Eve is able to gain about Alice's raw key. Note that this
information also depends on the measurements Eve performs on her part of the
states she shares with Bob. In general these will be positive operator-valued
measures (POVMs) which are allowed to depend on the basis (since we allow Eve
to possess a quantum memory). We call the POVM elements $F_{b0}$ and
$F_{b1}$, where $b \in \{0,1\}$ and $F_{b0} + F_{b1} = \id$. As will be
explained in the next subsection, we will be able to eliminate the explicit
appearance of the POVM elements in our optimisation problem.

\subsection{Eve's quantum error}
\label{sec:eves_error}

As stated in the introduction to this section, we wish to minimise the
extractable secret keyrate, which involves maximising the mutual information
$I(A : E)$. As a stepping stone to optimising this quantity we will consider
the QBER in Eve's inference of Alice's bits, $Q_{\mathrm{AE}}$, first
introduced in section~\ref{sec:intro}, in \eqref{eq:probs_q_delta}. Working
in a single basis $b$ for now, this quantity is given by
\begin{equation}
  Q^{(b)}_{\mathrm{AE}} = p^{(b)}_{\mathrm{AE}}(0,1)
  + p^{(b)}_{\mathrm{AE}}(1,0) \,.
\end{equation}
In general $I^{(b)}(A:E)$ depends on both this error $Q^{(b)}_{\mathrm{AE}}$
and the asymmetry $\delta^{(b)}$ also introduced in \eqref{eq:probs_q_delta},
and is an increasing function as $Q^{(b)}_{\mathrm{AE}}$ approaches 1/2 for
fixed $\delta^{(b)}$. Rather than attempting to \emph{directly} optimise the
mutual information in terms of $Q^{(b)}_{\mathrm{AE}}$ and $\delta^{(b)}$, we
instead turn our attention to the combination
\begin{equation}
  \label{eq:eve_weighted_error}
  Q^{(b)}_{\mathrm{AE}}(\varepsilon) = (1 + \varepsilon) p^{(b)}_{\mathrm{AE}}(0,1)
  + (1 - \varepsilon) p^{(b)}_{\mathrm{AE}}(1,0) \,.
\end{equation}
In terms of $Q^{(b)}_{\mathrm{AE}}$ and $\delta^{(b)}$ this is
\begin{equation}
  Q^{(b)}_{\mathrm{AE}}(\varepsilon) = Q^{(b)}_{\mathrm{AE}}
  + \varepsilon \delta^{(b)} \,.
\end{equation}
Optimising this quantity yields a $\delta^{(b)}$, dependent on the weighting
parameter $\varepsilon$, and an optimal $Q^{(b)}_{\mathrm{AE}}$ given
$\delta^{(b)}$. By varying $\varepsilon$ one may hope to sweep the range of
values of $\delta^{(b)}$ and obtain a profile of minimised
$Q^{(b)}_{\mathrm{AE}}$ as a function of $\delta^{(b)}$. The motivation for
this approach becomes apparent when we express
$Q^{(b)}_{\mathrm{AE}}(\varepsilon)$ in terms of Eve's probe and POVM
elements.

In terms of Eve's interaction and measurement,
\begin{subequations}
  \begin{IEEEeqnarray}{rCl}
    p^{(b)}_{\mathrm{AE}}(0,1)
    &=& \tfrac{1}{2} \Tr[\rho_{b0} F_{b1}] \,, \\
    p^{(b)}_{\mathrm{AE}}(1,0)
    &=& \tfrac{1}{2} \Tr[\rho_{b1} F_{b0}] \,,
  \end{IEEEeqnarray}
\end{subequations}
where $\rho_{bx} = \Tr_{\mathrm{B}}[\proj{\Psi_{bx}}]$, $\Tr_{\mathrm{B}}$ is
the partial trace over $\mathcal{H}_{\mathrm{B}}$, and $F_{bz}$ are POVM
elements which sum to unity for each basis. Substituting into
\eqref{eq:eve_weighted_error} and using that $F_{b1} = \id - F_{b0}$, we obtain
\begin{equation}
  Q^{(b)}_{\mathrm{AE}}(\varepsilon) = \tfrac{1}{2}(1 + \varepsilon)
  - \tfrac{1}{2} \Tr \bigl[ \bigl( (\rho_{b0} - \rho_{b1})
  + \varepsilon (\rho_{b0} + \rho_{b1}) \bigr) F_{b0} \bigr] \,.
\end{equation}
This expression is minimised by taking for $F_{b0}$ a projector which selects
the positive eigenvalue part of the operator in the trace (the Helstr\"om
bound). The result of optimising over Eve's measurement is
\begin{equation}
  Q^{(b)}_{\mathrm{AE}}(\varepsilon) = \tfrac{1}{2} - \tfrac{1}{4}
  \bnorm{(\rho_{b0} - \rho_{b1}) + \varepsilon(\rho_{b0} + \rho_{b1})}_{1} \,,
\end{equation}
where for an arbitrary matrix $\norm{M}_{1} = \Tr[(M^{\dagger}
M)^{1/2}]$. This replaces the explicit appearance of Eve's POVM with an
eigenvalue problem, leaving only an optimisation over Eve's interaction. Note
that this would not be possible if we instead attempted to optimise
$Q^{(b)}_{\mathrm{AE}}$ for fixed $\delta^{(b)}$, since in that case the POVM
element $F_{b0}$ would appear explicitly in the constraint as well as in the
expression to optimise.

Using $b = 0$ as an example, we now describe how we approach the problem of
maximising $Q^{(0)}_{\mathrm{AE}}$ and how we extract the corresponding
values of $Q^{(0)}_{\mathrm{AE}}$ and $\delta^{(0)}$. In terms of the four
states $\ket{a}$, $\ket{b}$, $\ket{c}$ and $\ket{d}$ introduced earlier in
order to parameterise the probe,
\begin{subequations}
  \begin{IEEEeqnarray}{rCl}
    \tfrac{1}{2} (\rho_{00} - \rho_{01})
    &=& \trans{a}{b} + \trans{b}{a} + \trans{c}{d} + \trans{d}{c} \,, \\
    \tfrac{1}{2} (\rho_{00} + \rho_{01})
    &=& \proj{a} + \proj{b} + \proj{c} + \proj{d} \,.
  \end{IEEEeqnarray}
\end{subequations}
In general our problem is to extract the eigenvalues of an operator $\hat{A}$
given its decomposition
\begin{equation}
  \hat{A} = A^{ij} \trans{i}{j}
\end{equation}
in terms of the states $\bigl\{\ket{i} \mid i \in \{a,b,c,d\}\bigr\}$ (where
we adopt the convention of summing over repeated indices). Explicitly
decomposing a vector $\ket{u}$ on the same basis as $\ket{u} = u^{i}
\ket{i}$, the action of $\hat{A}$ on $\ket{u}$ is
\begin{IEEEeqnarray}{rCl}
  \hat{A} \ket{u} &=& A^{ij} \trans{i}{j} u^{k} \ket{k} \IEEEnonumber \\
  &=& A^{ij} \gamma_{jk} u^{k} \ket{i} \,.
\end{IEEEeqnarray}
It is not difficult to see that determining the eigenvalues and eigenstates
of $\hat{A}$ is equivalent to determining the eigenvalues and eigenvectors of
the matrix $A \Gamma$, where $A = (A_{ij})$ and $\Gamma =
(\gamma_{ij})$. (This remains true even in the case where the vectors
$\{\ket{i}\}$ are not linearly independent.) The matrix whose eigenvalues we
wish to determine may be expressed as $D + \varepsilon \Gamma$, where
\begin{IEEEeqnarray}{rCl}
  D &=& \begin{bmatrix}
    \gamma_{ba} & b^{2} & \gamma_{bc} & \gamma_{dc} \\
    a^{2} & \gamma_{ab} & \gamma_{ac} & \gamma_{ad} \\
    \gamma_{da} & \gamma_{db} & \gamma_{dc} & d^{2} \\
    \gamma_{ca} & \gamma_{cb} & c^{2} & \gamma_{cd}
  \end{bmatrix} \,, \\
  \Gamma &=& \begin{bmatrix}
    a^{2} & \gamma_{ab} & \gamma_{ac} & \gamma_{ad} \\
    \gamma_{ba} & b^{2} & \gamma_{bc} & \gamma_{dc} \\
    \gamma_{ca} & \gamma_{cb} & c^{2} & \gamma_{cd} \\
    \gamma_{da} & \gamma_{db} & \gamma_{dc} & d^{2}
  \end{bmatrix} \,,
\end{IEEEeqnarray}
and $a^{2} = \gamma_{aa}$, and so on. Let the eigenvalues of this matrix be
$\{\lambda_{p}\}$ and the corresponding (not necessarily normalised)
eigenvectors be $\{v_{p}\}$, such that
\begin{equation}
  \label{eq:eig_problem}
  (D + \varepsilon \Gamma) v_{p} = \lambda_{p} v_{p} \,.
\end{equation}
In terms of the set of eigenvectors, the operator $F_{00}$ has the expression
\begin{equation}
  F_{00} = \sum_{\lambda_{p} > 0} \frac{\proj{v_{p}}}{\normsq{v_{p}}} \,,
\end{equation}
where $\ket{v_{p}} = v_{p}^{i} \ket{i}$, $i \in \{a,b,c,d\}$ and the sum is
over the indices $p$ for which $\lambda_{p} > 0$. Using this and that the
$\ket{v_{p}}$ are orthogonal, we obtain a matrix expression for the trace of
an arbitrary operator $\hat{A}$ multiplied by $F_{00}$:
\begin{IEEEeqnarray}{rCl}
  \Tr \bigl[ \hat{A} F_{00} \bigr]
  &=& \sum_{\lambda_{p} > 0} \frac{\bra{v_{p}} \hat{A} \ket{v_{p}}}
  {\normsq{v_{p}}} \IEEEnonumber \\
  &=& \sum_{\lambda_{p} > 0} \frac{v_{p}^{\dagger} \Gamma A \Gamma v_{p}}
  {v_{p}^{\dagger} \Gamma v_{p}} \,,
\end{IEEEeqnarray}
The explicit expressions for $Q^{(0)}_{\mathrm{AE}}$ and $\delta^{(0)}$ are
\begin{IEEEeqnarray}{rCl}
  Q^{(0)}_{\mathrm{AE}} &=& \tfrac{1}{2}
  - \sum_{\lambda_{p} > 0} \frac{v_{p}^{\dagger} \Gamma D v_{p}}
  {v_{p}^{\dagger} \Gamma v_{p}} \,, \\
  \delta^{(0)} &=& \tfrac{1}{2}
  - \sum_{\lambda_{p} > 0} \frac{v_{p}^{\dagger} \Gamma^{2} v_{p}}
  {v_{p}^{\dagger} \Gamma v_{p}} \,.
\end{IEEEeqnarray}
With $Q^{(0)}_{\mathrm{AE}}$ and $\delta^{(0)}$ determined, we have an
optimised value of $I^{(0)}(A:E)$ for fixed $\delta^{(0)}$, and all that
remains is to optimise $I^{(0)}(A:E)$ over $\varepsilon$.

Finally, the generalisation when we consider two bases is straightforward:
we will approach the optimisation of $I(A:E)$ by introducing three weighting
parameters $\varepsilon_{0}$, $\varepsilon_{1}$, and $\varepsilon$, instead
of one, optimising the quantity
\begin{equation}
  \label{eq:qasym2b}
  Q_{\mathrm{AE}}(\varepsilon_{0}, \varepsilon_{1}, \varepsilon)
  = \tfrac{1}{2}(1 + \varepsilon) Q^{(0)}_{\mathrm{AE}}(\varepsilon_{0})
  + \tfrac{1}{2}(1 - \varepsilon) Q^{(1)}_{\mathrm{AE}}(\varepsilon_{1}) \,,
\end{equation}
and then optimising $I(A:E)$ over $(\varepsilon_{0}, \varepsilon_{1},
\varepsilon)$.

\subsection{Inherent QBER}
\label{sec:inh_qber}

All that remains now, before being able to optimise \eqref{eq:qasym2b} over
all of Eve's possible unitary interactions, is to determine the full set of
constraints on the metric $\gamma_{ij}$, since not all metrics will represent
a unitary interaction, and to determine the relationship between the metrics
$\gamma_{ij}$ and $\gamma'_{ij}$ in the two bases (which depends only on the
angles $\theta$ and $\Delta$). This is done in the next subsection. Before
this, we demonstrate that there is a minimum nonzero QBER if $\alpha \neq
\beta$ (in which case Alice and Bob's bases cannot be perfectly
aligned). This is easily verified by expressing the QBER $Q$ in terms of a
basis $\{\ket{0'}, \ket{1'}\}$ intermediate between $\{\ket{\psX},
\ket{\psY}\}$ and $\{\ket{\psU}, \ket{\psV}\}$, and a basis $\{\ket{0},
\ket{1}\}$ midway between $\{\ket{\phx}, \ket{\phy}\}$ and $\{\ket{\phu},
\ket{\phv}\}$. Specifically,
\begin{subequations}
  \begin{IEEEeqnarray}{rCl}
    \ket{0'} &=& \cos\bigl(\tfrac{\alpha}{4}\bigr) \ket{\psX}
    + \sin\bigl(\tfrac{\alpha}{4}\bigr) \ket{\psY} \,, \\
    \ket{1'} &=& \cos\bigl(\tfrac{\alpha}{4}\bigr) \ket{\psY}
    - \sin\bigl(\tfrac{\alpha}{4}\bigr) \ket{\psX} \,,
  \end{IEEEeqnarray}
\end{subequations}
and
\begin{subequations}
  \begin{IEEEeqnarray}{rCl}
    \ket{0} &=& \cos\bigl(\tfrac{\beta}{4}\bigr) \ket{\phx}
    + \sin\bigl(\tfrac{\beta}{4}\bigr) \ket{\phy} \,, \\
    \ket{1} &=& \cos\bigl(\tfrac{\beta}{4}\bigr) \ket{\phy}
    - \sin\bigl(\tfrac{\beta}{4}\bigr) \ket{\phx} \,.
  \end{IEEEeqnarray}
\end{subequations}

Setting
\begin{subequations}
  \begin{IEEEeqnarray}{rCl}
    \Sigma_{z} &=& \proj{0'} - \proj{1'} \,, \\
    \Sigma_{x} &=& \trans{0'}{1'} + \trans{1'}{0'} \,,
  \end{IEEEeqnarray}
\end{subequations}
and
\begin{subequations}
  \begin{IEEEeqnarray}{rCl}
    \sigma_{z} &=& \proj{0} - \proj{1} \,, \\
    \sigma_{x} &=& \trans{0}{1} + \trans{1}{0} \,,
  \end{IEEEeqnarray}
\end{subequations}
then with this choice of basis the expression we find for the quantum error
is
\begin{IEEEeqnarray}{rCl}
  Q &=& \tfrac{1}{2} - \tfrac{1}{4} \cos\bigl(\tfrac{\alpha}{2}\bigr)
  \cos\bigl(\tfrac{\beta}{2}\bigr)
  \Tr \bigl[ \Sigma_{z} (\sigma_{z} \otimes \id_{\mathrm{E}}) \bigr]
  \IEEEnonumber \\
  && - \tfrac{1}{4} \sin\bigl(\tfrac{\alpha}{2}\bigr)
  \sin\bigl(\tfrac{\beta}{2}\bigr)
  \Tr \bigl[ \Sigma_{x} (\sigma_{x} \otimes \id_{\mathrm{E}}) \bigr] \,.
\end{IEEEeqnarray}
Clearly, $-2 \leq \Tr \bigl[ \Sigma_{z} (\sigma_{z} \otimes \id_{\mathrm{E}})
\bigr] \leq 2$ and $-2 \leq \Tr \bigl[ \Sigma_{x} (\sigma_{x} \otimes
\id_{\mathrm{E}}) \bigr] \leq 2$, and we find the bound
\begin{equation}
  \label{eq:lower_q_ae}
  Q \geq \tfrac{1}{2} - \tfrac{1}{2}
  \max\{\abs{\cos(\Delta)}, \abs{\cos(\theta)}\} \,,
\end{equation}
with $\Delta$ and $\theta$ defined as in \eqref{eq:defs_delta_theta} (this
bound is also saturated, e.g.\ if Eve does not interfere with the channel, in
which case $\Sigma_{z,x} = \sigma_{z,x}$). The corresponding upper bound is
\begin{equation}
  \label{eq:upper_q_ae}
  Q \leq \tfrac{1}{2} + \tfrac{1}{2}
  \max\{\abs{\cos(\Delta)}, \abs{\cos(\theta)}\} \,.
\end{equation}

\subsection{Transformation and constraints}
\label{sec:diff_constraints}

We now determine the full set of constraints on the metric elements
$\gamma_{ij}$. First, we impose that the QBER is fixed at $Q$. This, combined
with $\braket{\psX}{\psX} = \braket{\psY}{\psY} = 1$, imposes
\begin{subequations}
  \begin{IEEEeqnarray}{rCl}
    a^{2} + b^{2} &=& 1 - Q \,, \\
    c^{2} + d^{2} &=& Q \,,
  \end{IEEEeqnarray}
\end{subequations}
and $\re[\gamma_{ab}] = \re[\gamma_{cd}] = 0$, with analogous constraints for
the basis $b = 1$. The components $\gamma_{ab}$, $\gamma_{ac}$,
$\gamma_{bd}$, and $\gamma_{cd}$ transform between the two bases according to
\begin{subequations}
  \begin{IEEEeqnarray}{rCl}
    \label{eq:transf_g_ab}
    \gamma'_{ab} &=& \cos(\Delta) \cos(\theta) \gamma_{ab}
    + \cos(\Delta) \sin(\theta) \gamma_{ac} \IEEEnonumber \\
    &&+ \sin(\Delta) \cos(\theta) \gamma_{db}
    + \sin(\Delta) \cos(\theta) \gamma_{dc} \,, \\
    \gamma'_{ac} &=& \cos(\Delta) \cos(\theta) \gamma_{ac}
    - \cos(\Delta) \sin(\theta) \gamma_{ab} \IEEEnonumber \\
    &&+ \sin(\Delta) \cos(\theta) \gamma_{dc}
    - \sin(\Delta) \sin(\theta) \gamma_{db} \,, \\
    \gamma'_{db} &=& \cos(\Delta) \cos(\theta) \gamma_{db}
    + \cos(\Delta) \sin(\theta) \gamma_{dc} \IEEEnonumber \\
    &&- \sin(\Delta) \cos(\theta) \gamma_{ab}
    - \sin(\Delta) \sin(\theta) \gamma_{ac} \,, \\
    \label{eq:transf_g_dc}   
    \gamma'_{dc} &=& \cos(\Delta) \cos(\theta) \gamma_{dc}
    - \cos(\Delta) \sin(\theta) \gamma_{db} \IEEEnonumber \\
    &&- \sin(\Delta) \cos(\theta) \gamma_{ac}
    + \sin(\Delta) \sin(\theta) \gamma_{ab} \,.
  \end{IEEEeqnarray}
\end{subequations}
For a more compact representation, the transformation matrix from
$[\gamma_{ab},\, \gamma_{ac},\, \gamma_{db},\, \gamma_{dc}]^{T}$ to
$[\gamma'_{ab},\, \gamma'_{ac},\, \gamma'_{db},\, \gamma'_{dc}]^{T}$ can be
expressed as
\begin{equation}
  \begin{bmatrix}
    \phantom{-} \cos(\Delta) & \sin(\Delta) \\
    - \sin(\Delta) & \cos(\Delta)
  \end{bmatrix}
  \otimes
  \begin{bmatrix}
    \phantom{-} \cos(\theta) & \sin(\theta) \\
    - \sin(\theta) & \cos(\theta)
  \end{bmatrix} \,.
\end{equation}
\eqref{eq:transf_g_ab} and \eqref{eq:transf_g_dc} together with the
constraint $\re[\gamma_{ab}] = \re[\gamma_{cd}] = 0$ imply $\re[\gamma_{ac}]
= \re[\gamma_{bd}] = 0$.

For $a'$ and $d'$, we find
\begin{subequations}
  \label{eq:transf_ad}
  \begin{IEEEeqnarray}{rCl}
    a'^{2} &=& \cos(\Delta)^{2} a^{2} + \sin(\Delta)^{2} d^{2}
    + \sin(2\Delta) \re[\gamma_{ad}] \,, \\
    d'^{2} &=& \cos(\Delta)^{2} d^{2} + \sin(\Delta)^{2} a^{2}
    - \sin(2\Delta) \re[\gamma_{ad}] \,,
  \end{IEEEeqnarray}
\end{subequations}
from which we immediately see that $a'^{2} + d'^{2} = a^{2} + d^{2}$. From
\eqref{eq:transf_ad}, and taking the real and imaginary parts of
\begin{equation}
  \gamma'_{ad} = - \tfrac{1}{2} \sin(2\Delta) (a^{2} - d^{2})
  + \cos(\Delta)^{2} \gamma_{ad} - \sin(\Delta)^{2} \gamma_{da} \,,
\end{equation}
we find
\begin{subequations}
  \begin{IEEEeqnarray}{rCl}
    \delta'_{ad} &=& \cos(2\Delta) \delta_{ad}
    + \sin(2\Delta) \re[\gamma_{ad}] \,, \\
    \re[\gamma'_{ad}] &=& \cos(2\Delta) \re[\gamma_{ad}]
    - \sin(2\Delta) \delta_{ad} \,, \\
    \im[\gamma'_{ad}] &=& \im[\gamma_{ad}] \,,
  \end{IEEEeqnarray}
\end{subequations}
where $\delta_{ad} = \frac{a^{2}-d^{2}}{2}$. Similarly, $b'^{2} + c'^{2} =
b^{2} + c^{2}$ and
\begin{subequations}
  \begin{IEEEeqnarray}{rCl}
    \delta'_{bc} &=& \cos(2\theta) \delta_{bc}
    + \sin(2\theta) \re[\gamma_{bc}] \,, \\
    \re[\gamma'_{bc}] &=& \cos(2\theta) \re[\gamma_{bc}]
    - \sin(2\theta) \delta_{bc} \,, \\
    \im[\gamma'_{bc}] &=& \im[\gamma_{bc}] \,,
  \end{IEEEeqnarray}
\end{subequations}
with $\delta_{bc} = \frac{b^{2}-c^{2}}{2}$. Orthogonality of $\ket{\psX}$ and
$\ket{\psY}$ implies $\im[\gamma_{bc}] = \im[\gamma_{ad}]$.

We still require $a'^{2} \leq 1-Q$ and $d'^{2} \leq Q$ individually, which
impose
\begin{subequations}
  \label{eq:constr_ad_prime}
  \begin{IEEEeqnarray}{rCl}
     \label{eq:constr_a_prime}
    \cos(\Delta)^{2} a^{2} + \sin(\Delta)^{2} d^{2}
    + \sin(2\Delta) \re[\gamma_{ad}] &\leq& 1 - Q \,, \\
    \label{eq:constr_d_prime}
    \cos(\Delta)^{2} d^{2} + \sin(\Delta)^{2} a^{2}
    - \sin(2\Delta) \re[\gamma_{ad}] &\leq& Q \,.
  \end{IEEEeqnarray}
\end{subequations}
Equation~\eqref{eq:constr_a_prime} is automatically satisfied, in the sense
that there are no new restrictions on $a^{2}$, $d^{2}$, or
$\re[\gamma_{ad}]$, if $Q \geq \tfrac{1}{2} - \tfrac{1}{2}
\abs{\cos(\Delta)}$. Equation~\eqref{eq:constr_d_prime} is automatically
satisfied if $Q \leq \tfrac{1}{2} + \tfrac{1}{2}
\abs{\cos(\Delta)}$. Similarly, we automatically have $b'^{2} \leq 1 - Q$ and
$c'^{2} \leq Q$ as long as $\tfrac{1}{2} - \tfrac{1}{2} \abs{\cos(\theta)}
\leq Q \leq \tfrac{1}{2} + \tfrac{1}{2} \abs{\cos(\theta)}$.

Finally, using $a'^{2} + b'^{2} = a^{2} + b^{2}$ and $c'^{2} + d'^{2} = c^{2}
+ d^{2}$, we obtain the constraint
\begin{equation}
  \label{eq:constr_ad_bc}
  \sin(2\Delta) \re[\gamma_{ad}] + \sin(2\theta) \re[\gamma_{bc}]
  = \sin(\Delta)^{2}(a^{2} - d^{2}) + \sin(\theta)^{2}(b^{2} - c^{2}) \,.
\end{equation}

\subsection{Optimisation}

The plots given in figures~\ref{fig:keyrate_qber} and \ref{fig:keyrate_theta}
were generated by numerically maximising $Q_{\mathrm{AE}} =
Q_{\mathrm{AE}}(\varepsilon_{0} = \varepsilon_{1} = \varepsilon = 0)$,
defined by equation \eqref{eq:qasym2b}, using MATLAB's \verb|fmincon|
routine, over all metrics $\gamma_{ij}$ respecting the constraints derived in
the preceding subsection for the reported angles $\theta$ and values of
$Q_{\mathrm{AB}}$ and with $\Delta = 0$, and calculating the corresponding
value of $I(A:E)$. For simplicity, we performed no systematic optimisation
over $(\varepsilon_{0}, \varepsilon_{1}, \varepsilon)$. Optimising over
$(\varepsilon_{0}, \varepsilon_{1}, \varepsilon)$ in a few test cases
generally supported our expectation that the minimal keyrate would be
obtained for the maximal value of $Q_{\mathrm{AE}}$ with a symmetric attack
($\delta^{(0)} = \delta^{(1)} = 0$ and $Q^{(0)}_{\mathrm{AE}} =
Q^{(1)}_{\mathrm{AE}}$). Similarly, investigating test cases generally found
that the minimal keyrate, given a common error bound on the deviation of
$\alpha$ and $\beta$ from $90^{\circ}$, was obtained by setting both to the
worst case such that $\alpha = \beta = \theta$ and $\Delta = 0$. As a result,
the keyrates given in section~\ref{sec:opt_results} are an upper bound on the
secure keyrate (which is sufficient to demonstrate a degradation in
performance) which we believe are very likely the optimal keyrates.

The maximum tolerable QBERs reported in figure~\ref{fig:secure_qber} are
those for which $Q = Q_{\mathrm{AE}}$ for the angles $\theta$ considered,
again with $\Delta = 0$.

In addition to the keyrates reported in section~\ref{sec:opt_results}, we
also similarly investigated the case in which only one basis is used to
generate the key, by maximising only $Q^{(0)}_{\mathrm{AE}}$. In this case,
the resulting keyrates (not accounting for sifting) are lower than those
obtained for the case in which both bases are used, for the same
parameters. This suggests that implementations of BB84 in which both bases
are used to generate the key are likely to be more robust against
implementation errors, as we alluded to in section~\ref{sec:discussion}.

\section{Acknowledgements}

This work was supported by the European EU QCS project, the CHIST-ERA DIQIP
project, the Interuniversity Attraction Poles Photonics@be Programme (Belgian
Science Policy), and the Brussels-Capital Region through a BB2B Grant. Erik
Woodhead acknowledges support from the Belgian Fonds pour la Formation \`a la
Recherche dans l'Industrie et dans l'Agriculture (F.R.I.A.).

\bibliography{qkd_errors}

\end{document}